\documentclass[aps,prl,twocolumn,superscriptaddress]{revtex4}

\usepackage[compat=1.0.0]{tikz-feynman}
\usepackage{amssymb}
\usepackage{multirow}
\usepackage{bm}
\usepackage{amsmath}
\usepackage{graphicx}
\usepackage{epstopdf}
\usepackage{subfigure}
\usepackage{natbib}
\usepackage{epsfig}
\usepackage{amsfonts}
\usepackage{mathrsfs}
\usepackage{braket}
\usepackage[toc,page,title,titletoc,header]{appendix}
\usepackage[colorlinks,linkcolor=blue,citecolor=blue,anchorcolor=blue]{hyperref}
\usepackage{dsfont,amsthm,amsbsy}

\usepackage{fancyhdr}

\usetikzlibrary{quotes,angles}
\newcommand{\obtuseangle}{\kern.08em
\begin{tikzpicture}
    \draw coordinate (a) at (0.14,0);
    \draw coordinate (b) at (0,0);
    \draw coordinate (c) at (-.12,0.18);
    \draw (a) -- (b) -- (c) pic [draw=black]{} ;
\end{tikzpicture}%
\kern.08em%
}

\begin{document}
\title{Stripe order enhanced superconductivity in the Hubbard model}
\author{Hong-Chen Jiang}
\affiliation{Stanford Institute for Materials and Energy Sciences, SLAC National Accelerator Laboratory and Stanford University, Menlo Park, California 94025, USA}
\author{Steven A. Kivelson}
 \affiliation{Department of Physics, Stanford University, Stanford, California 94305, USA}

\date{\today}
\begin{abstract}
Unidirectional (``stripe'') charge-density-wave order has now been established as a ubiquitous feature in the phase diagram of the cuprate high temperature (HT) superconductors, where it generally competes with superconductivity (SC).  None-the-less, on theoretical grounds it has  been conjectured that stripe order (or other forms of ``optimal'' inhomogeneities) may play an essential positive role in the mechanism of HTSC.  Here we report density matrix renormalization group studies of the Hubbard model on long 4 and 6 leg cylinders where the hopping matrix elements transverse to the long direction are periodically modulated - mimicing the effect of putative period-2 stripe order.  We find even modest amplitude modulations can enhance the long-distance SC correlations by many orders of magnitude, and drive the system into a phase with a substantial spin gap and SC quasi-long-range-order with a Luttinger exponent, $K_{sc} \sim 1$.
\end{abstract}
\maketitle

A complex relation between multiple ordering tendencies appears to be a universal feature of highly correlated electronic systems\cite{tranquadarmp}. For example, charge-density-wave (CDW), spin-density-wave (SDW), and d-wave superconducting (SC) orders all arise in significantly overlapping regimes of the phase diagram of the cuprate high temperature superconductors.  Moreover, studies of the Hubbard model with repulsive $U$ of order the band-width, $U\sim W$, have been difficult to interpret unambiguously, in large part because these same ordering tendencies appear to be in delicate balance with one another\cite{arovasreview,gullreview}.

There are clear senses in which these orders ``compete'':  This can be seen phenomenologically in the cuprates where suppressing SC order with a magnetic field enhances the strength of the observed CDW, and where the most robust SC often appears in regions of the phase diagram where the CDW order is relatively weaker\cite{review}. A similar feature is vividly apparent in density matrix renormalization group (DMRG) studies of the  the Hubbard model on long but relatively narrow cylinders and 
ladders\cite{competition,scalapinowhitereview,2jeckelmanstripes,Jiang2018tJ,Dodaro2017,Jiang2019Hub,Jiang2020Hub,whitesimons,whitenew,Qin2020,White2021,Jiang2020tJ,Gong2021}. Here, the closest possible approximation of a SC state is a Luther-Emery liquid,\cite{lutheremery} in which the SC and CDW susceptibilities are determined by quantum mechanically dual variables. Thus,  any change in the parameters - e.g. details of the band-structure or the strength of the interactions -   that  enhances the long distance correlations of one necessarily  decreases the other.  It has even been suggested that this competition is so ferocious that  the Hubbard model with $U\sim W$ may  never be superconducting in the 2d limit.\cite{Qin2020}

However, the fact that high temperature superconductivity and CDW (not to mention SDW) orders all seem to appear together suggests that they may be linked in a more multifacetted manner than the word ``competing'' suggests.\cite{ineluctable}  Indeed, two distinct theoretical proposals carry the implication   that CDW order can enhance SC: 1) It was proposed in Ref.\cite{italians1,italians2} that CDW fluctuations - associated with proximity to a putative CDW quantum critical point - could serve as an effective pairing ``glue'' and thereby enhance SC even under conditions in which fully developed CDW order might depress SC by openning gaps on portions of the Fermi surface.  2) It was proposed in Ref. \cite{emerymezachar}, and further developed in a variety of subsequent papers\cite{arrigoni,optimalinhom,scalettarcheckerboard,dror1,dror2,yaocheckerboard}, that static or slowly fluctuating CDW order could produce a form of ``optimally inhomogeneous'' electronic structure that could enhance SC.

\begin{figure}
  \includegraphics[width=0.8\linewidth]{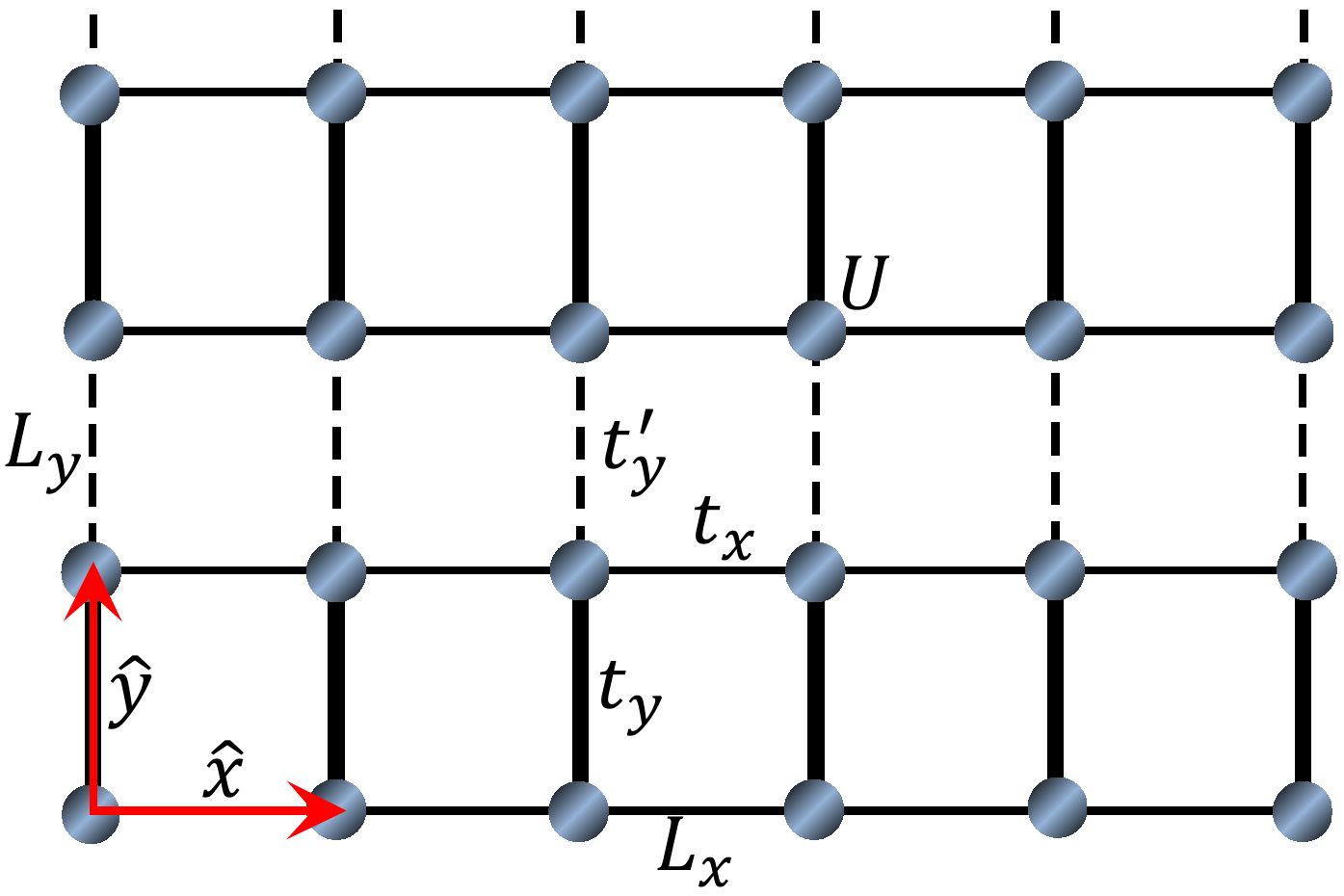}
  \caption{(Color online) Hubbard model on the square cylinder. Periodic and open boundary conditions are imposed, respectively, along the directions specified by the lattice basis vectors $\hat{y}=(0,1)$ and $\hat{x}=(1,0)$. $t_x=t$ and $t_y=t+dt$ ($t_y^\prime = t-dt$) are hopping integrals between nearest-neighbor sites in the $\hat{x}$ and $\hat{y}$ directions. $U$ is the on-site Coulomb repulsion, $L_x$ and $L_y$ are the number of sites.}\label{Fig:Lattice}
\end{figure}

In the present paper, we use DMRG studies of the square lattice Hubbard model on 4 and 6 leg cylinders with length $L_x=32$ and $48$ to explore the second of these propositions. We consider the model with only nearest-neighbor interactions $t$, with $U=12t$, and for electron density per site $n=1-\delta$ with $\delta = 1/8$ and $1/12$. Moreover, we assume an ordered period 2 explicit CDW with ordering vector perpendicular to the long axis of the cylinder, so that the hopping-matrix elements in this direction are alternately enhanced or depressed, $ t\to t\pm dt$ as shown in Fig.\ref{Fig:Lattice}.  

For $dt=0$ this is the uniform Hubbard model, which in this range of parameters appears\cite{Dodaro2017,Jiang2019Hub,Jiang2020Hub,Qin2020,White2021} to favor an insulating  phase with spontaneous translation symmetry breaking corresponding to an array of ``full stripes,'' i.e. the CDW period along the cylinder is $\lambda_{cdw} = 1/\delta$.\cite{Zaanen1989} As might be expected, this state has exponentially falling SC correlations at long distances. For $dt=t$, this system consists of decoupled 2-leg ladders. While the behavior of the 2-leg ladder depends on the ratio of $t_y/t_x$, so long as this ratio does not exceed a critical value\cite{Jiang2020tJ}, the 2-leg ladder is  known\cite{noak2leg,White2002,troyer,arrigoni,Jiang2020tJ} to support a Luther-Emery liquid phase with power law SC correlations that fall with distance $r$ as $|r|^{-K_{sc}}$ with $K_{sc} $ between 1 and 2.\footnote{Note that in the model as defined, the decoupled 2-leg ladder limit reached when $dt\to t$ has $t_y/t_x = 2$, which exceeds the critical value at which the Luther-Emery phase is observed;  however, since this limit could be approached in multiple ways, the intuition that the finite $dt$ state can be thought of from the perspective of weakly coupled Luther-Emery liquids is probably still valid.} Here, we explore the effect of relatively weak modulations, $dt\leq 0.4$. 

In all  cases we find that the modulation enhances the SC correlations at long distances relative to the uniform cylinder ($dt=0$) by many orders of magnitude.  Indeed,  the  modulated cylinder seemingly forms a Luther-Emery liquid: The spin-spin correlator and the single-particle Green function fall exponentially with distance with a correlation length of order a lattice constant, indicating the existence of a spin gap.  Moreover, there are clear CDW correlations with wavelength $\lambda_{cdw} = 1/2\delta$ for the 4 leg and  $\lambda_{cdw} = 2/3\delta$ for the 6 leg cylinder. However, while it is plausible that they also have power-law correlations characterized by Luttinger exponent $K_{cdw}$, the expected duality relation $K_{cdw} = 1/K_{sc}$ is only barely consistent with the DMRG results for the 4-leg and clearly inconsistent with them for the 6-leg cylinder. Thus, unambiguous identification of the conformal field theory that characterizes the long-distance properties of the 6-leg cylinder is still a work in progress.

%====Fig.1====

{\bf The model:}
We employ DMRG\cite{White1992} to study the ground state properties of the Hubbard model on the square lattice, which is defined by the Hamiltonian%
\begin{eqnarray}\label{Eq:Ham}
H=-\sum_{\langle ij\rangle\sigma} t_{ij} \left(\hat{c}^\dagger_{i\sigma} \hat{c}_{j\sigma} + h.c.\right) + U\sum_i\hat{n}_{i\uparrow}\hat{n}_{i\downarrow}.
\end{eqnarray}
Here $\hat{c}^\dagger_{i\sigma}$ ($\hat{c}_{i\sigma}$) is the electron creation (annihilation) operator on site $i=(x_i,y_i)$ with spin polarization $\sigma$, and $\hat{n}_{i\sigma}$ is the electron number operator. We take the lattice geometry to be cylindrical with periodic (open) boundary condition in the $\hat{y}$ ($\hat{x}$) direction, as shown in Fig.\ref{Fig:Lattice}. $\langle ij\rangle$ denotes nearest-neighbor (NN) sites. $t_x=t$, $t_y=t+dt$, and $t_y^\prime=t-dt$ are the electron hopping integrals between NN sites in the $\hat{x}$ and $\hat{y}$ directions, respectively. Here, we focus on cylinders with width $L_y$ and length $L_x$, where $L_x$ and $L_y$ are the number of sites along the $\hat{x}$ and $\hat{y}$ directions, respectively. The total number of sites is $N=L_x\times L_y$, the number of electrons is $N_e$, and the doping level of the system is defined as $\delta=N_h/N$, where $N_h=N-N_e$ is the number of doped holes relative to the half-filled insulator that arises when $N_e=N$.

In the present study, we chose units of energy such that $t=1$ and consider $dt\leq 0.4$. We consider $U=12$ at $\delta=1/12$ and $\delta=1/8$ doping levels and focus on $L_y=4$ and $6$ leg cylinders of length up to $L_x=48$. We perform around 60 sweeps and keep up to $m=20000$ number of states for $L_y=4$ cylinders with a typical truncation error $\epsilon\sim 5\times 10^{-7}$, and up to $m=35000$ states for $L_y=6$ cylinders with a typical truncation error $\epsilon\sim 3\times 10^{-6}$.

%==Fig2: Superconductivity==
\begin{figure}
  \includegraphics[width=\linewidth]{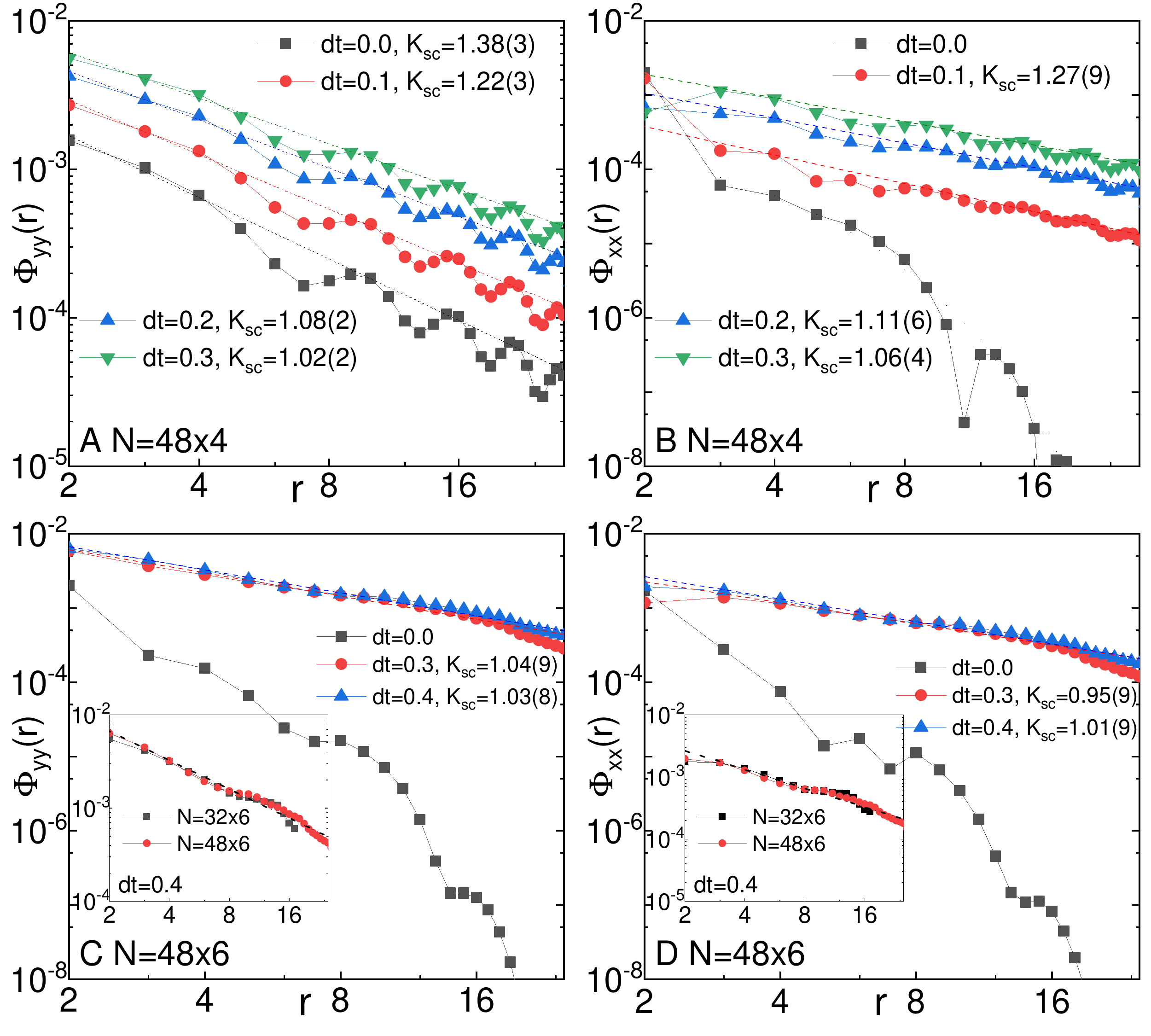}
  \caption{(Color online) Superconducting pair-field correlations. (A) $\Phi_{yy}(r;1,0)$ and (B) $\Phi_{xx}(r;1,0)$ on $N=48\times 4$ cylinders at $\delta=1/12$ with different $dt$, (C) $\Phi_{yy}(r;1,0)$ and (D) $\Phi_{xx}(r;1,0)$ on $N=48\times 6$ cylinders at $\delta=1/12$ with different $dt$ on double-logarithmic scales. Insets: $\Phi_{yy}(r;1,0)$ and $\Phi_{xx}(r;1,0)$ in double-logarithmic scales with $dt=0.4$ on both $N=32\times 6$ and $N=48\times 6$ cylinders. $r$ is the distance between two Cooper pairs in the $\hat{x}$ direction. Note that only the central-half region with $2\leq r\leq L_x/2+1$ is shown and used in the fitting, whereas the remaining data points from each end are removed to minimize boundary effects. The dashed lines denote power-law fitting to $\Phi(r)\sim r^{-K_{sc}}$.}\label{Fig:SC}
\end{figure}

The results of our calculations (as explained below) are summarized for $\delta=1/12$ in the remaining figures and quantified in Table \ref{table}.  More details, including further analysis of truncation error and results for $\delta=1/8$, are provided in the Supplemental Material (SM).

\begin{table}[]
    \centering
    \begin{tabular}{|c|c||c|c|c|c||c|c|c|}
    %\begin{tabular}{|M{1.5cm}|M{1.5cm}| M{1.5cm}|M{1.5cm}|M{1.5cm}|M{1.5cm}|M{1.5cm}|M{1.5cm}|M{1.5cm}|}
    \hline
     $L_y$ & $dt$ & $K_{sc}$  & $\Delta_d$ & $\Delta_s$ & $\Delta_\pi$ & $K_{cdw}$ & $\xi_{s}$ & $\xi_G$ \\
     \hline
     \hline
     4 &  0.0 & 1.38(3) & 0.0 & 0.0 & {0.066} & 1.27(1) & 8.6(4) &3.9(2) \\
     \hline
     4 &  0.1 & 1.22(3) &  {0.019} & {-0.011} & {0.074} & 1.35(1) & 7.1(2) &3.6(2) \\
   \hline
     4 &  0.2 & 1.08(2) &  {0.032} & {-0.016} & {0.082} & 1.46(1) & 4.7(2) &3.0(1) \\
     \hline
     4 &  0.3 & 1.02(2) &  {0.042} & {-0.021} & {0.091} & 1.48(1) & 2.9(1) &2.5(1) \\
      \hline
     \hline
     6 &  0.0 & $\infty$ & 0.0 & 0.0 & 0.0 & 0.3(3) & {3.9(4)} & {2.4(3)} \\
     \hline
     6 &  0.3 & 1.04(9) &  {0.070} & {0.004} & {0.038} & 3.5(2) & 1.7(1) &1.8(1) \\
     \hline
     6 &  0.4 & 1.03(8) &  {0.062} & {-0.011} & {0.065} & 3.3(2) & 1.3(1) &2.2(1) \\
     \hline
    \end{tabular}
    \caption{Extracted parameters obtained by fitting the DMRG results to theoretically expected asymptotic forms of various correlation functions for $\delta=1/12$ and the given values of  $L_y$ and $dt$.  Exponentially falling correlations are represented by a Luttinger exponent of $\infty$.  Precise levels of uncertainty due to finite size effects -- especially with regard to the Luttinger exponents -- are difficult to estimate.}\label{table}
\end{table}

{\bf Superconducting pair-field correlations: }%
We have calculated  the equal-time spin-singlet  SC pair-field correlation function %
\begin{eqnarray}\label{Eq:SC}
\Phi_{\alpha\beta}(r;y_0,y)=\langle\Delta^{\dagger}_{\alpha}(x_0,y_0)\Delta_{\beta}(x_0+r,y_0+y)\rangle.
\end{eqnarray}
Here
$\Delta^{\dagger}_{\alpha}(x,y)=\frac{1}{\sqrt{2}}[\hat{c}^{\dagger}_{(x,y),\uparrow}\hat{c}^{\dagger}_{(x,y)+\alpha,\downarrow}+\hat{c}^{\dagger}_{(x,y)+\alpha,\uparrow}\hat{c}^{\dagger}_{(x,y),\downarrow}]$ is the spin-singlet pair creation operator on bond $\alpha=\hat{x}$ or $\hat{y}$,  ($x_0,y_0$) identifies a site chosen with $x_0\sim L_x/4$ and $r$ ($y$) is the  displacement between bonds in the $\hat x$ ($\hat y$) direction. At long distances ($r \gg 1$), $\Phi_{\alpha \beta}$ exhibits power-law decay -- i.e quasi long-range order (QLRO) -- characterized by the Luttinger exponent $K_{sc}$:
\begin{eqnarray}
\Phi_{\alpha \beta}(r;y_0,y)\sim r^{-K_{sc}}\ \Delta_\alpha(y_0)\ \Delta_\beta(y_0+y)\label{Eq:Ksc}
\end{eqnarray}

The spatial symmetries of the striped model are such that there are two inequivalent y-directed bonds and a unique x directed bond. Consistent with this, there are three distinct pair-field amplitudes, which we identify with distinct symmetries of pairing as they arise in the limit $dt=0$ and $L_y\to \infty$:
\begin{align}
  & \Delta_{y}(y) = \Delta_s+\Delta_d + e^{i\pi (y-1)} \Delta_\pi \\
  & \Delta_{x}(y) = \Delta_s-\Delta_d \nonumber 
\end{align}
Since for finite $L_y$ there is no exact symmetry that exchanges the x and y axes, there is no sharp distinction between d-wave and (extended) s-wave order, but it is reasonable (and conventional) to refer to the case in which $|\Delta_d|$ is the largest component as ``d-wave-like'' pairing.  For $dt=0$, there is a sharp distinction between ``$\pi$ pairing'' ($\Delta_\pi\neq 0$ and $\Delta_d=\Delta_s=0$) and d-wave like pairing (with $\Delta_\pi=0$ and $\Delta_d\neq 0$);  since for $L_y=4$, $\pi$ pairing  is equivalent to d-wave pairing on plaquettes oriented perpendicular to the long axis of the cylinder, such a state has been referred to in this context as ``true d-wave''\cite{Dodaro2017} or ``plaquette d-wave''\cite{Chung2020} pairing.  For non-zero $dt$, by symmetry we would expect all (or none) of these components to be present, but we can loosely identify distinct states by which component is largest (dominant). (These symmetry arguments are made more precise in Sec. D in SM.)

Fig.\ref{Fig:SC}A shows $\Phi_{yy}(r;1,0)$, i.e. between $t_y$ bonds, for $L_y=4$ cylinders at $\delta=1/12$. The exponent $K_{sc}$, obtained by fitting the results using Eq.(\ref{Eq:Ksc}), is $K_{sc}=1.38(3)$ for the uniform case, $dt=0.0$, while for  $dt=0.2 - 0.3$, $K_{sc}\sim 1$. We have also computed other components of $\Phi_{\alpha \beta}$:  $\Phi_{xx}(r;1,0)$ is shown in Fig.\ref{Fig:SC}B and $\Phi_{xy}(r;1,0)$ and $\Phi_{yy}(r;1,1)$ are shown in Fig.S2 in the SM. For the isotropic case with $dt=0.0$, $\Phi_{xx}(r;1,0)$ and $\Phi_{xy}(r;1,0)$ decay exponentially as $\Phi_{xx}(r;1,0)\sim e^{-r/\xi_{sc}}$ with $\xi_{sc}\sim 1.8$,\cite{Jiang2018tJ,Chung2020} and $\Phi_{yy}(r;1,y) \sim (-1)^y$, i.e. the amplitudes are consistent with $\pi$-pairing QLRO with {$\Delta_\pi = 0.066$} and $\Delta_d=\Delta_s=0$. This is consistent with previous studies of the $L_y=4$ Hubbard and $t$-$J$ models with $dt=0$.\cite{Jiang2018tJ,Jiang2019Hub,Jiang2020Hub,Chung2020} 
The key new observation is that $\Phi_{xx}(r;y_0,0)$ and $\Phi_{xy}(r;y_0,0)$ are significantly enhanced for $dt>0$, so that they decay as a power-law with a {similar} $K_{sc}$ as $\Phi_{yy}$. In particular, not only is $K_{sc}$ decreased from its $dt=0$ value, $|\Delta_d|$ increases rapidly  as well. For example, for $dt=0.3$, $\Delta_d=0.042$, $\Delta_s=-0.021$ and $\Delta_\pi=0.091$. (More complete results are presented in Table \ref{table}.)

%==Fig3: Ni and Kc==
\begin{figure}
  \includegraphics[width=\linewidth]{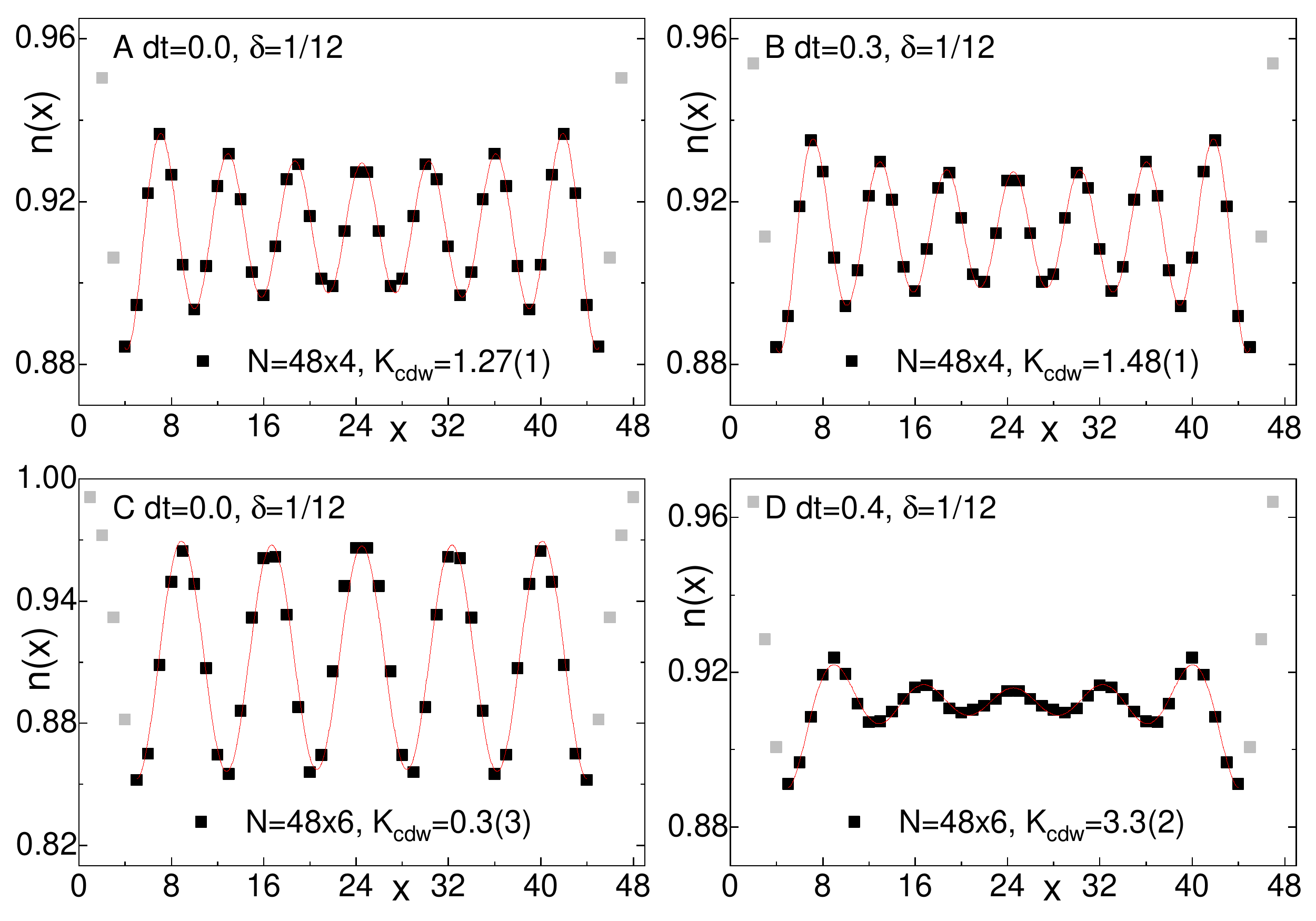}
  \caption{(Color online) Charge density profiles. Charge density distribution $n(x)$ at $\delta=1/12$ doping level on $N=48\times 4$ cylinders with (A) $dt=0.0$ and (B) $dt=0.3$, and on $N=48\times 6$ cylinders with (C) $dt=0.0$ and (D) $dt=0.4$. The exponent $K_{cdw}$ is extracted using Eq.(\ref{Eq:Kc}) where the red lines are fitting curves. A few data points in light grey are neglected to minimize boundary effects.}\label{Fig:CDW}
\end{figure}

The results are still more dramatic for $L_y=6$:  Consistent with previous studies on the isotropic Hubbard model, on $L_y=6$ cylinders with $dt=0$ we find the SC correlations are relatively weak and appear to decay exponentially with distance as shown, for $\delta = 1/12$, in Fig.\ref{Fig:SC} C and D. However, as was the case for $L_y=4$ cylinders, we find that the SC pair-field correlations are dramatically enhanced by a finite $dt>0$, where we find that $\Phi_{\alpha\beta}(r) \sim r^{-K_{sc}}$ with $K_{sc}\sim 1$. Moreover, the SC pairing symmetry is d-wave like with $\Phi_{xx}(r) \sim \Phi_{yy}(r) \sim -\Phi_{xy}(r)$. For example, for $dt=0.3$, {$\Delta_d=0.042$, $\Delta_s=0.004$ and $\Delta_\pi = 0.038$}. As summarized in the SM, the results we have obtained for $\delta=1/8$ are qualitatively similar to those with $\delta=1/12$. For instance, for $dt=0.3$ at $\delta=1/8$, $K_{sc}=1.07(7)$, $\Delta_d=0.074$, $\Delta_s=0.007$ and $\Delta_\pi=0.032$.

%==Fig4: Spin-spin correlation==
\begin{figure}
  \includegraphics[width=\linewidth]{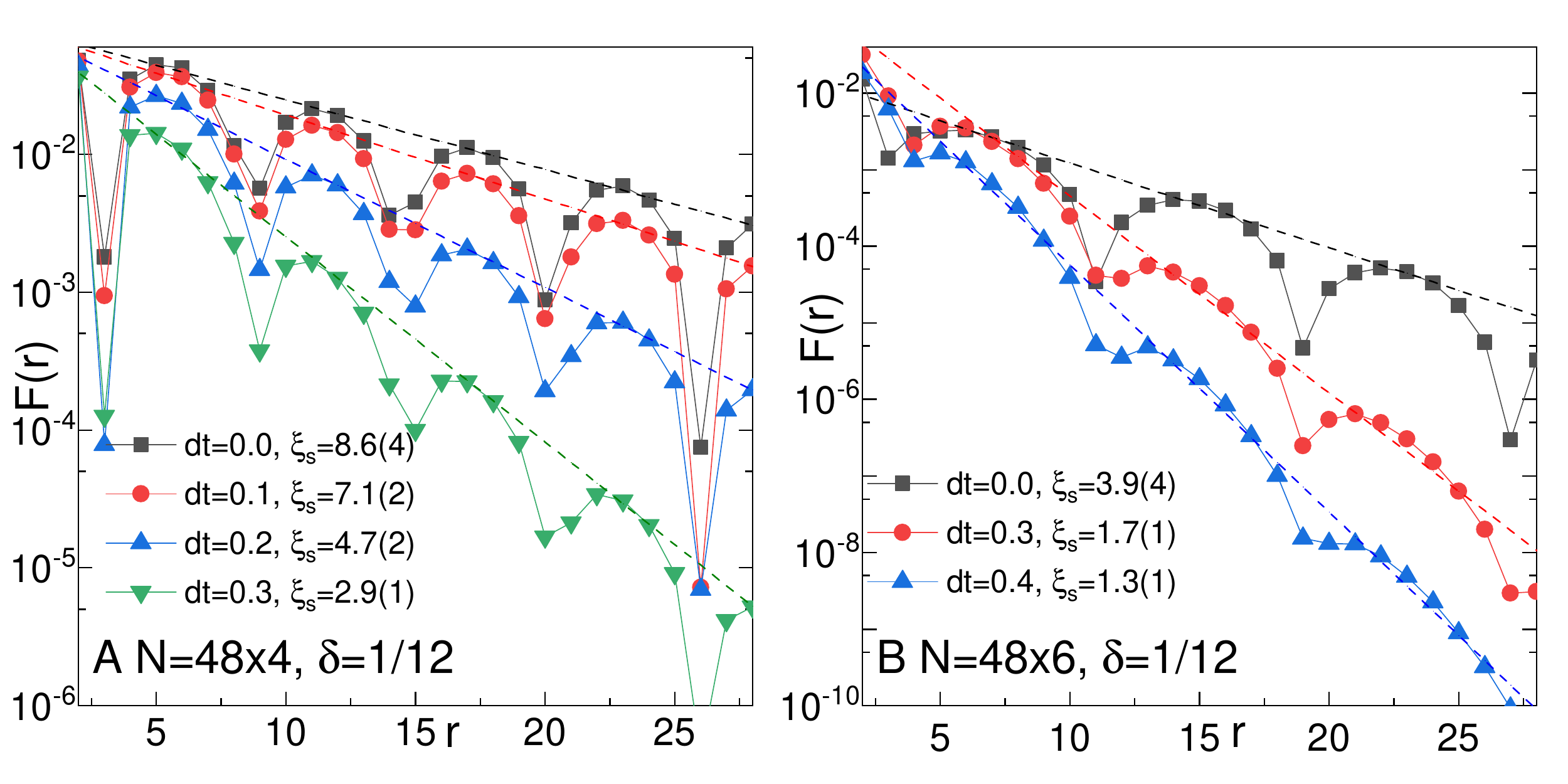}
  \caption{(Color online) Spin-spin correlations at $\delta=1/12$. (A) $F(r)$ on $N=48\times 4$ cylinders with different $dt$, and (B) $F(r)$ on $N=48\times 6$ cylinders with different $dt$, in semi-logarithmic scale. Dashed lines denote exponential fit $F(r)\sim e^{-r/\xi_s}$, where $r$ is the distance between two sites in the $\hat{x}$ direction.}\label{Fig:SpinCor}
\end{figure}

{\bf CDW correlations: }%
To measure the charge order, we define the rung density operator $\hat{n}(x)=L_y^{-1}\sum_{y=1}^{L_y}\hat{n}(x,y)$ and its expectation value $n(x)=\langle \hat{n}(x)\rangle$. Fig.\ref{Fig:CDW}A-B shows the charge density distribution $n(x)$ for $L_y=4$ cylinders, which is consistent with ``half-filled charge stripes'' with  wavelength $\lambda_{cdw}=1/2\delta$. This corresponds to an ordering wavevector $Q=4\pi\delta$, i.e. viewing the cylinder as a 1D system, 2 holes per 1D unit cell. The charge density profile $n(x)$ for $L_y=6$ cylinders is shown in Fig.\ref{Fig:CDW}C-D, which has wavelength $\lambda_{cdw}=2/3\delta$, consistent with ``two-third-filled" charge stripes. This corresponds to an ordering wavevector $Q=3\pi\delta$, i.e. 4 holes per 1D unit cell.

At long distance, the spatial decay of the CDW correlation is dominated by a power-law with the Luttinger exponent $K_{cdw}$. The exponent $K_{cdw}$ can be obtained by fitting the charge density oscillations induced by the boundaries of the cylinder\cite{White2002,Gong2021}
\begin{eqnarray}\label{Eq:Kc}
n(x)&=&n_0 + A(x)* \cos(Qx+\phi) \\
A(x)&=&A_Q*(x^{-K_{cdw}/2}+(L_x+1-x)^{-K_{cdw}/2}).\nonumber
\end{eqnarray}
Here $A_Q$ is an amplitude, $\phi$ is a phase shift,  $n_0=1-\delta$ is the mean density, and $Q=4\pi\delta$ for $L_y=4$ cylinders and $Q=3\pi \delta$ for $L_y=6$ cylinders.  Note that to improve the fitting quality, a few data points (corresponding to the light grey points Fig.\ref{Fig:CDW}) are excluded to minimize the boundary effect. Values of $K_{cdw}$ are summarized in Table \ref{table}. The fact that $K_{cdw}> K_{sc}$ for all cases in which $dt>0$ suggests that CDW order is secondary compared with SC. The one exception is $L_y=6$ and $dt=0$, where the CDW correlations are at best slowly decaying and are clearly stronger than the SC. Our results are consistent with {CDW  QLRO with a value of $K_{cdw}\leq 0.3$,} consistent with previous results for the $t$-$J$ model.\cite{Qin2020} Note that similar values of $K_{cdw}$ can also be obtained from the asymptotic fall-off of the density-density correlation function, as shown in the SM.

%===Fig.5 Single-particle correlation===
\begin{figure}[tb]
\centering
    \includegraphics[width=1\linewidth]{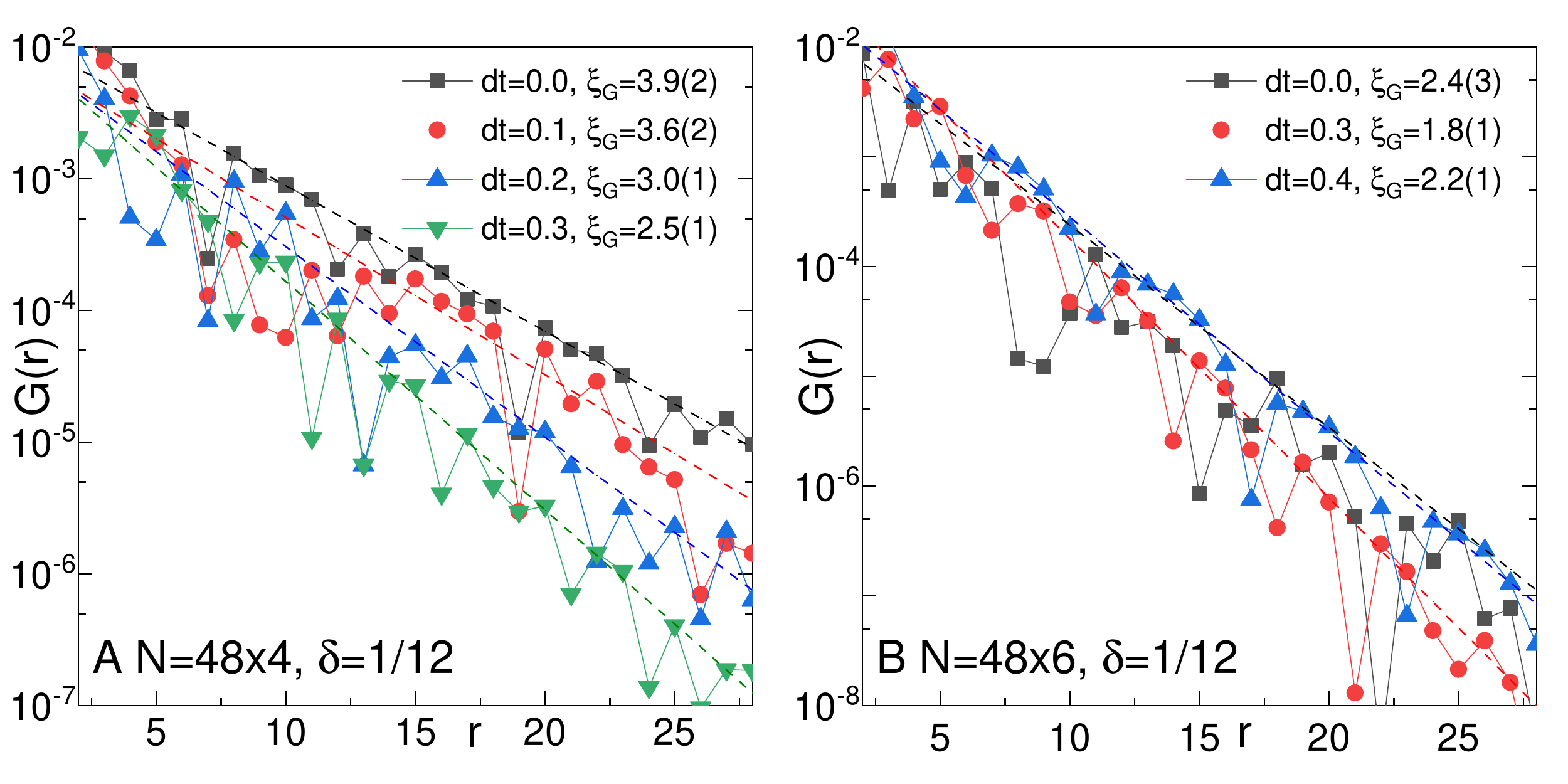}
\caption{(Color online) Single-particle Green function at $\delta=1/12$. (A) $G(r)$ on $N=48\times 4$ cylinders with different $dt$, and (B) $G(r)$ on $N=48\times 6$ cylinders with different $dt$ on the semi-logarithmic scale. Dashed lines denote exponential fitting $G(r)\sim e^{-r/\xi_G}$ where $r$ is the distance between two sites in the $\hat{x}$ direction.}\label{Fig:CC}
\end{figure}

{\bf Spin-spin correlations: }%
To describe the magnetic properties of the ground state, we calculate the spin-spin correlation functions defined as%
\begin{eqnarray}\label{Eq:SpinCor}
F(r)=\langle \vec{S}_{x_0,y_0}\cdot \vec{S}_{x_0+r,y_0}\rangle. 
\end{eqnarray}
Here $\vec{S}_{x,y}$ is the spin operator on site $i=(x,y)$ and $i_0=(x_0,y_0)$ is the reference site with $x_0\sim L_x/4$. Fig.\ref{Fig:SpinCor} shows $F(r)$ for both $L_y=4$ and $L_y=6$ cylinders at $\delta=1/12$ with different $dt$. It is clear that $F(r)$ decays exponentially as $F(r)\sim e^{-r/\xi_s}$ at long-distances, with a finite correlation length $\xi_s$, i.e. there must be a finite gap in the spin sector.  Moreover, $\xi_s$ decreases with  increasing $dt$ on both $L_y=4$ and $L_y=6$ cylinders. In addition, we also observe for both $L_y=4$ and $L_y=6$ cylinders that the spin-spin correlation has spatial modulation with a wavelength $\lambda_s$ that is twice that of the charge, i.e., $\lambda_s=2\lambda_{cdw}$. Values of $\xi_s$ for $\delta=1/12$ and various values of $dt$ are given in Table \ref{table}.

{\bf Single particle Green function: }%
We have also calculated the single-particle Green function, defined as%
\begin{eqnarray}\label{Eq:CC}
G(r)=\langle c^{\dagger}_{(x_0,y),\sigma} c_{(x_0+r,y),\sigma}\rangle.
\end{eqnarray}
Fig.\ref{Fig:CC} shows  $G(r)$ for both $L_y=4$ and $L_y=6$ cylinders at $\delta=1/12$ with different $dt$. The long distance behavior of $G(r)$ is consistent with exponential
decay $G(r)\sim e^{-r/\xi_G}$. The extracted correlation lengths $\xi_G<4$ for both $L_y=4$ and $L_y=6$ cylinders are comparable to $\xi_s$, as also shown in Table \ref{table}.

{\bf Summary of Results:}
What we have generically found, both for $L_y=4$ and $L_y=6$, over the entire investigated range of stripe modulation strength, $dt$, and  doped hole concentration, $\delta$, is a form of SC QLRO with exponentially falling spin and single particle correlations and with typically weaker, but presumably also power-law correlated CDW QLRO.  Expressed in terms of the various quantities extracted by the above discussed fits of the DMRG results to theoretically expected asymptotic forms are summarized (without error bars) in Table \ref{table}.

{\bf Conclusions:}
It is both conceptually and practically important to understand what aspects of electronic structure are optimal for superconductivity.  Circumstantial evidence has been adduced in several ways that certain organized forms of spatially inhomogeneous structure can enhance superconductivity, but we feel that the present results constitute the clearest and most unambiguous evidence to date that this is a real and robust effect.  More generally, concerning the question of whether the 2d Hubbard model can support high temperature superconductivity - the present results offer encouraging evidence of an affirmative answer, as they  constitute some of the strongest long-range superconducting correlations documented to date on systems wider than 4 legs.  It is worth acknowledging that the present results on period 2 CDW order cannot be directly compared with the situation in the cuprates, where  the CDW order  typically has period closer to 3 (YBCO) or 4 (BSCCO and LSCO).  None-the-less, it suggests that a more nuanced approach to the intertwining of CDW and SC orders may be appropriate in the cuprate context. 

Finally, there is the question of obtaining a conceptual understanding of the numerical results we have reported.  This is an ongoing endeavor.  However, it is worth mentioning a possible connection between the present results, and recent DMRG results that exhibit enhanced superconductivity in a lightly doped quantum spin liquid.\cite{Jiang2021Square}  Indeed, in the discussion of the  ``spin-gap proximity effect''  in Ref. \cite{emerymezachar}, an analogy was made between the  effects of stripe order and a mechanism based on a doped spin liquid. 

It is reasonable to conclude that the low energy (gapless) magnetic fluctuations associated with  antiferromagneitc order or near order, are determinantal to SC - they would generally be expected to be pair-breaking.  However, higher energy, short-range correlated antiferrommagnetic fluctuations can produce precisely the sort of momentum dependent interactions that are most condusive to d-wave SC.  In this sense, a fully gapped spin liquid would seem to have just the right spectrum of magnetic fluctuations to be an optimal parent to a high temperature supercondcutor.  Indeed, it is possible to view the gap in such a state as the pairing gap of a superconductor that is waiting to be liberated.  In a similar sense, the undoped ($\delta=0$) %$n=1$)
two-leg Hubbard ladder has a spin-gap and can be viewed as a Mott insulator of preexisting Cooper pairs (rung singlets).  In this sense, doping into a modulated array of effective two leg ladders may be analogous to doping a fully gapped quantum spin liquid.

%==Acknowledgements==
{\it Acknowledgments:} We would like to thank Dror Orgad, Vladimir Calvera, Richard Scalettar, Doug Scalapino, John Tranquada and Thomas Devereaux for helpful discussions. This work was supported by the Department of Energy, Office of Science, Basic Energy Sciences, Materials Sciences and Engineering Division, under Contract DE-AC02-76SF00515.

%\bibliographystyle{unsrt}
%\bibliography{Refs}

%%%%%%%%%%%%%%%%
\clearpage
%\appendix 
\newpage

\renewcommand{\thefigure}{S\arabic{figure}}
\setcounter{figure}{0}%reset counter
\renewcommand{\theequation}{S\arabic{equation}}%redefine %command that creates equation no.
\setcounter{equation}{0}%reset counter
\setcounter{page}{1}
\renewcommand{\thetable}{S\arabic{table}}
\setcounter{table}{0}
%\makeatletter

\begin{center}
\noindent {\large {\bf Supplemental Material}}
\end{center}

%==Numerical convergence==
\section{A. More numerical details}\label{SM:Detail} %
We have checked the numerical convergence of our DMRG simulations by testing various symmetries of the results, such as spin rotational symmetry. The ground state of a finite system cannot spontaneously break any continuous symmetry. Therefore, the true ground state of the Hubbard model on a finite cylinder should preserve the $SU(2)$ spin rotational symmetry. This is a key indicator that  a DMRG simulation has converged to the true ground state. 

We take two routes to address this issue in our DMRG simulations. Firstly, we calculate the expectation value of the $z$-component of the spin operators $\langle\hat{S}^z_i\rangle$, which should be zero on any lattice site $i$ in the true ground state. Indeed, on the $L_y=4$ cylinders with different $dt$ we find that $\langle \hat{S}^z_i\rangle=0$ for all $i$ when we keep a number of states $m\geq 4096$, while on the $L_y=6$ cylinders with $dt=0.3$ and $dt=0.4$ the same is true when we keep a number of states $m\geq 8000$. Moreover, the spin $SU(2)$ symmetry requires that the relation $\langle S^x_i S^x_j\rangle$=$\langle S^y_i S^y_j\rangle$=$\langle S^z_i S^z_j\rangle$ holds for any pair of sites $i$ and $j$, which is again fulfilled in our simulations. In addition to spin rotational symmetry, we have verified that other symmetries including both the lattice translational symmetry in the $\hat{y}$ direction and reflection symmetry in the $\hat{x}$ direction are also satisfied. 

That all these symmetries are respected is strong evidence that our results are converged. We have also explored the effect of cylinder size and boundary effects. That nothing significant changes upon going from $L_x=32$ to $L_x=48$ suggests  that our systems are long enough to be extrapolated to the $L_x\to \infty$ limit.

%==Superconducting correlation==
\section{B. Superconducting correlations}\label{SM:SC} %
\begin{figure}
  \includegraphics[width=\linewidth]{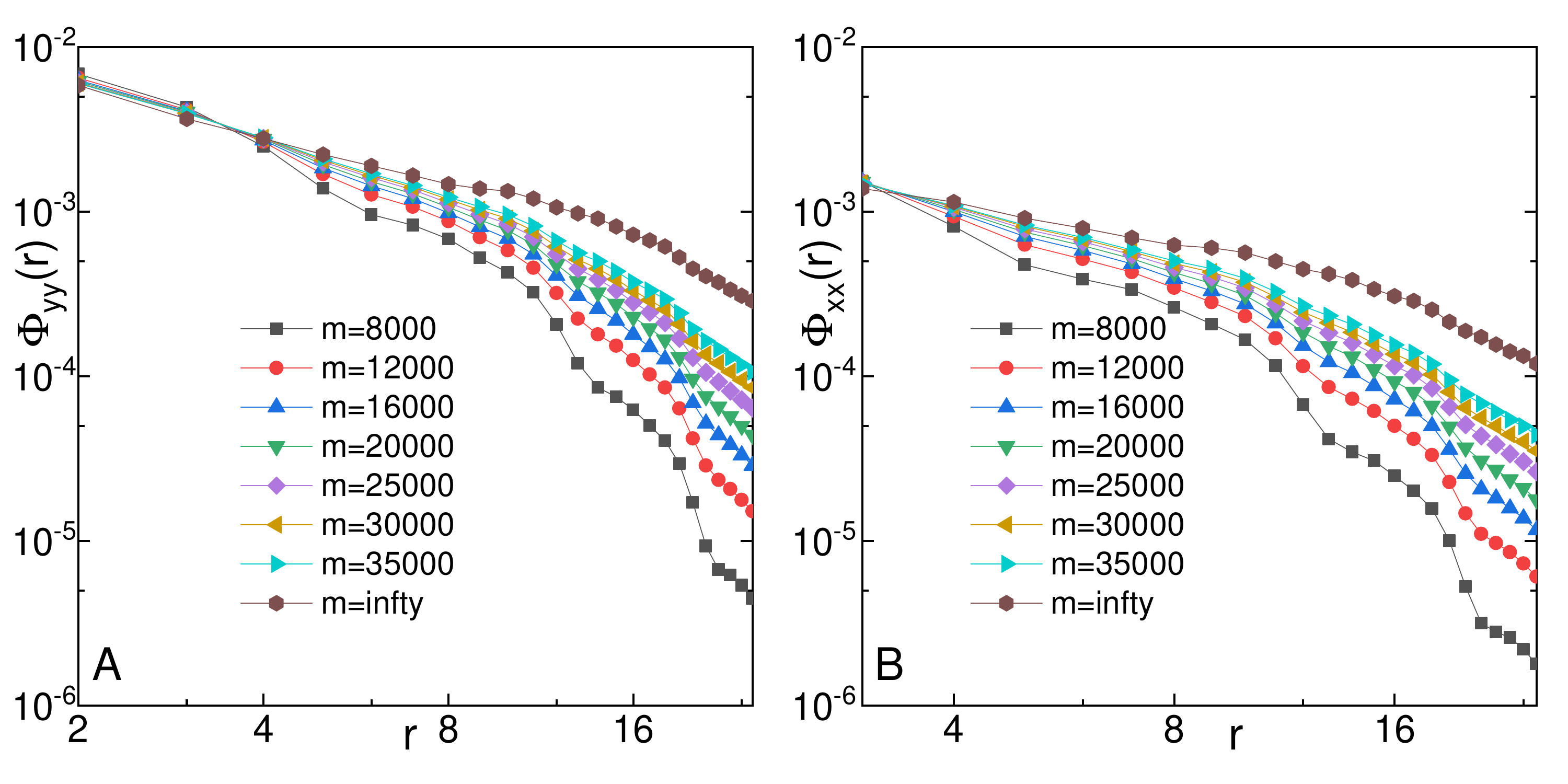}
  \caption{Convergence of superconducting correlations. SC correlation (A) $\Phi_{yy}(r;1,0)$ and (B) $\Phi_{xx})(r;1,0)$ on $N=48\times 6$ cylinder at $\delta=1/12$, by keeping $m$ number of states and its extrapolation in the limit $m=\infty$ on double-logarithmic scales. Here $r$ is the distance between two Cooper pairs in the $\hat{x}$ direction. Note that only the central-half region with $2\leq r\leq L_x/2+1$ is shown and used in the fitting, whereas the remaining data points from each end are removed to minimize boundary effects.}\label{FigS:SC}
\end{figure}

Fig.\ref{FigS:SC} shows the superconducting (SC) pair-field correlations $\Phi_{yy}(r;1,0)$ and $\Phi_{xx}(r;1,0)$ with $y_0=1$ for $N=48\times 6$ cylinders at $\delta=1/12$ and $dt=0.4$. The extrapolated $\Phi_{yy}(r;1,0)$ and $\Phi_{xx}(r;1,0)$ in the limit $m=\infty$ or $\epsilon=0$ is obtained by using a second-order polynomial function to fit the four data points with the largest number of states. To minimize the boundary and finite-size effects, the first few data points with small $r$ are excluded. As indicated by the red dashed lines, the SC correlations are consistent with a power-law decay $\Phi(r)\propto r^{-K_{sc}}$ with $K_{sc}\approx 1$. We have used the same procedure for both $L_y=4$ and $L_y=6$ cylinders and obtained the extrapolated $\Phi(r)$ in the limit $m=\infty$ or $\epsilon=0$. In addition to the spin-singlet SC correlations, we have also calculated the spin-triplet SC correlations. However, these are much weaker than the spin-singlet SC correlation, suggesting that spin-triplet superconductivity is unlikely.

\begin{figure}
  \includegraphics[width=\linewidth]{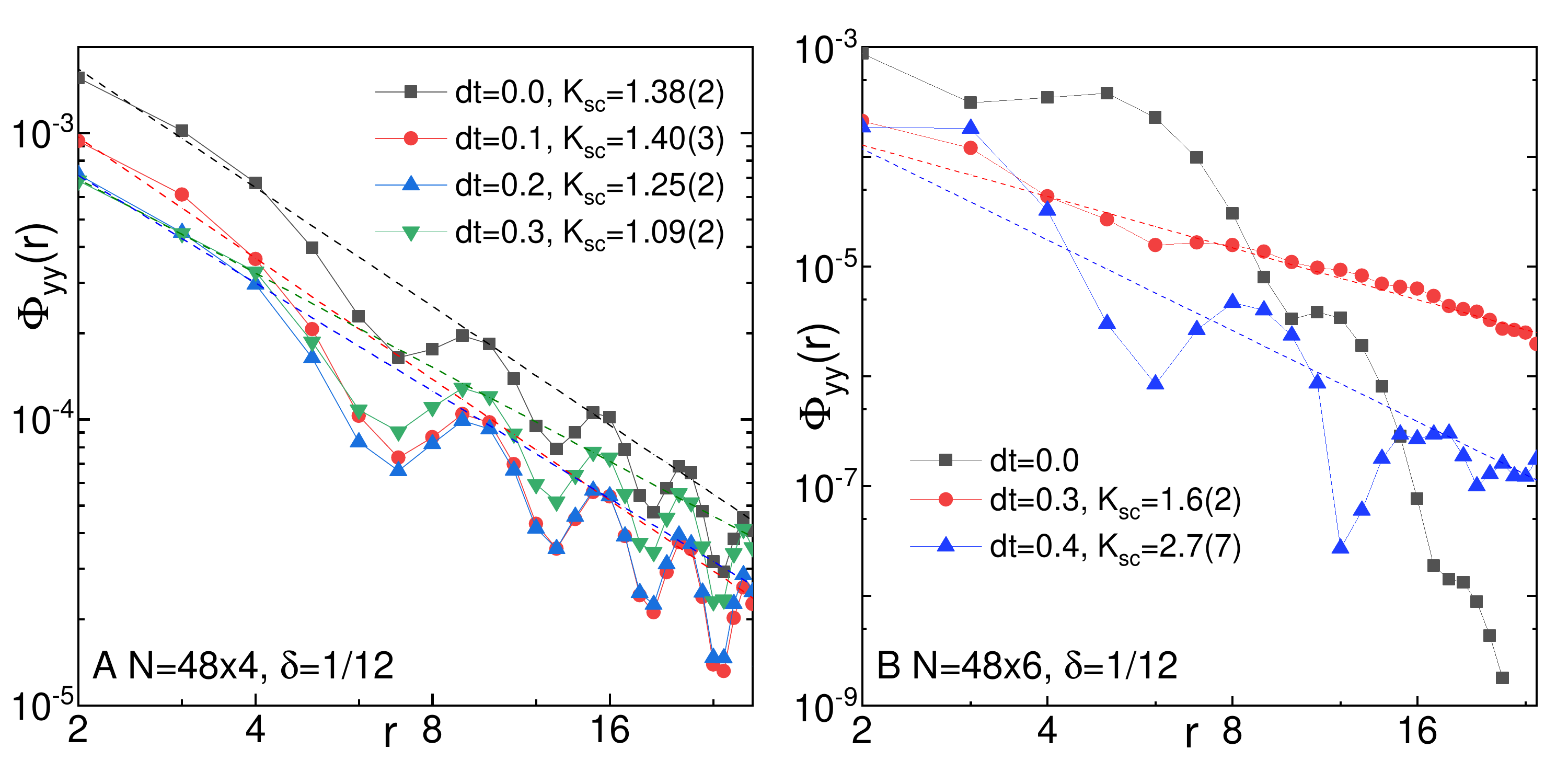}
  \caption{Superconducting correlations. $\Phi_{yy}(r;2,0)$ between $t_y^\prime$ bonds at $\delta=1/12$ on (A) $N=48\times 4$ and (B) $N=48\times 6$ cylinder at different $dt$ on double-logarithmic scales. Here $r$ is the distance between two Cooper pairs in the $\hat{x}$ direction. Dashed lines label the power-law fit $\Phi_{yy}(r)\sim 1/r^{K_{sc}}$.}\label{FigS:SCtyp}
\end{figure}

Fig.\ref{FigS:SCtyp} shows the SC correlations, $\Phi_{yy}(r;2,0)$, i.e., between $t_y^\prime$ bonds, for both $L_y=4$ and $L_y=6$ cylinders at $\delta=1/12$ and different $dt$'s. For $L_y=4$ cylinders with $dt=0.2 - 0.3$, the SC correlations are consistent with a power-law decay $\Phi(r)\sim r^{-K_{sc}}$ with an exponent $K_{sc}\approx 1$. For $L_y=6$ cylinders, while the SC correlations are also consistent with a power-law decay with an exponent $K_{sc}=1.6(2)$ for $dt=0.3$ and $K_{sc}=2.7(7)$ for $dt=0.4$, they are notably weaker than $\Phi_{yy}(r;1,0)$, i.e., between $t_y$ bonds, shown in the main text. 

%==FigS1: Superconducting correlation==

%==FigS2: Superconducting correlation==

%==FigS3: Superconducting correlation==
\begin{figure}
  \includegraphics[width=\linewidth]{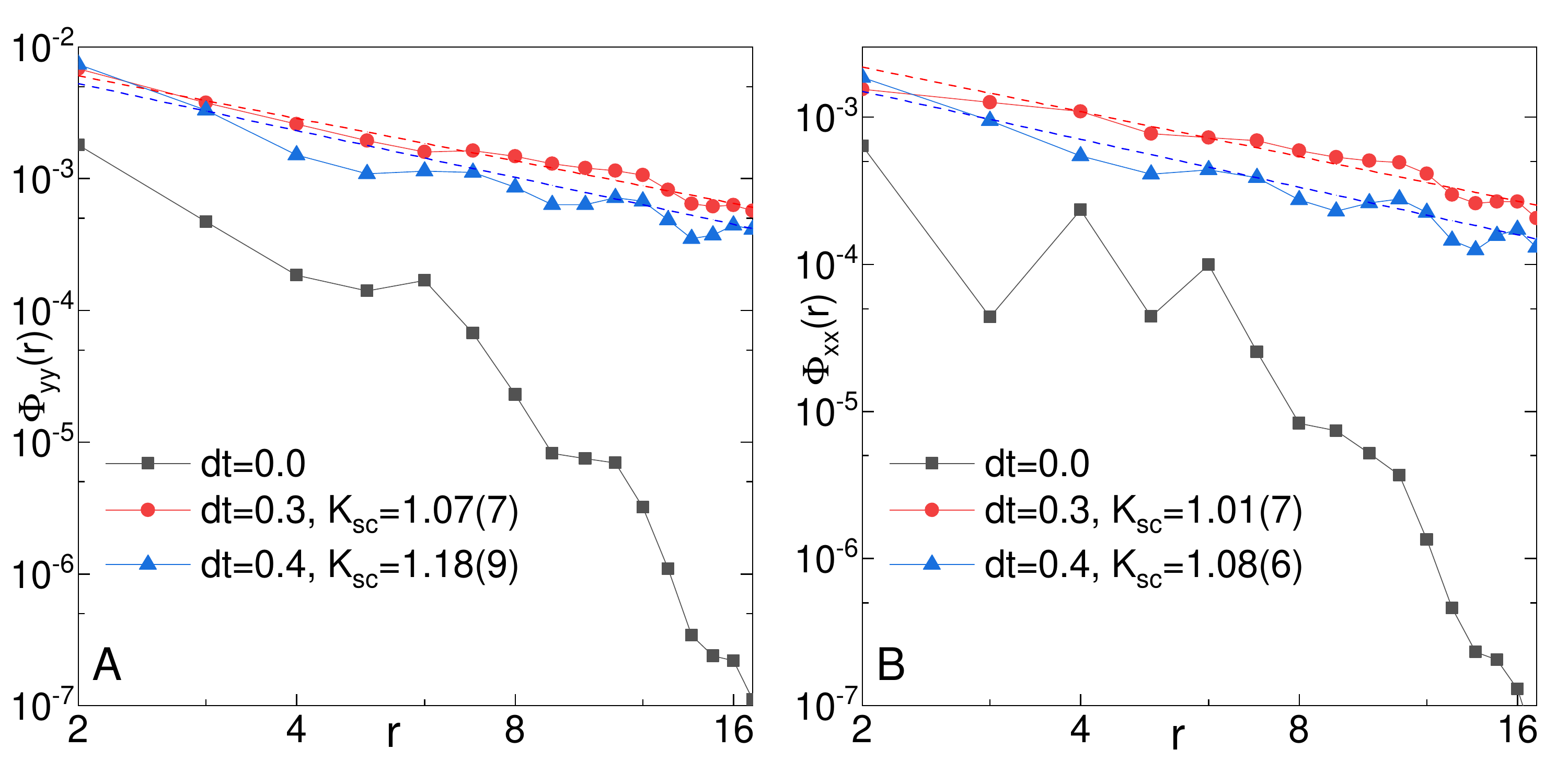}
  \caption{Superconducting correlations. (A) $\Phi_{yy}(r;1,0)$ and (B) $\Phi_{xx}(r;1,0)$ on $N=32\times 6$ cylinder at $\delta=1/8$ with different $dt$ on double-logarithmic cales. Here $r$ is the distance between two Cooper pairs in the $\hat{x}$ direction. Note that only the central-half region with $2\leq r\leq L_x/2+1$ is shown and used in the fitting, whereas the remaining data points from each end are removed to minimize boundary effects. Dashed lines label the power-law fit $\Phi(r)\sim 1/r^{K_{sc}}$.}\label{FigS:SCR8All}
\end{figure}

\begin{table}[]
    \centering
    \begin{tabular}{|c||c|c|c|c||c|c|c|c|}
    %\begin{tabular}{|M{1.5cm}|M{1.5cm}| M{1.5cm}|M{1.5cm}|M{1.5cm}|M{1.5cm}|M{1.5cm}|M{1.5cm}|M{1.5cm}|}
    \hline
     $dt$ & $K_{sc}$  & $\Delta_d$ & $\Delta_s$ & $\Delta_\pi$ & $K_{cdw}$ & $\xi_{s}$ & $\xi_G$ \\
     \hline
     \hline
     0.0 & $\infty$ & 0.0 & 0.0 & 0.0 & 0.6(2) &  3.8(6)  & 2.6(4) \\
     \hline
     0.3 & 1.07(7) &  {0.074} & {0.007} & {0.032} & 3.7(7) & 1.1(1) &2.4(1) \\
     \hline
     0.4 & 1.18(9) &  {0.059} & {0.003} & {0.047} & 1.8(3) & 1.5(1) &3.3(2) \\
     \hline
    \end{tabular}
    \caption{Extracted parameters obtained by fitting the DMRG results to theoretically expected asymptotic forms of various correlation functions for $\delta=1/8$ and the given values of $dt$ on $N=32\times 6$ cylinder.  Exponentially falling correlations are represented by a Luttinger exponent of $\infty$.  Precise levels of uncertainty due to finite size effects -- especially with regard to the Luttinger exponents -- are difficult to estimate.}\label{Stable}
\end{table}

Fig.\ref{FigS:SCR8All} shows $\Phi_{yy}(r;1,0)$, i.e. between $t_y$ bonds, and $\Phi_{xx}(r;1,0)$, i.e., between $t_x$ bonds, for $N=32\times 6$ cylinder at $\delta=1/8$. Consistent with previous studies on the isotropic Hubbard model with $dt=0$, we find the SC correlations are relatively weak and appear to decay exponentially with distance $r$ and $K_{sc}=\infty$. However, our results show that the SC correlations are dramatically enhanced by a finite $dt$, for instance, $dt=0.3$ and $dt=0.4$, where we find that $\Phi_{\alpha\beta}(r)\sim r^{-K_{sc}}$ with $K_{sc}\sim 1$. Similar with the case at $\delta=1/12$, for $dt=0$, $\Delta_d=\Delta_s=\Delta_\pi=0$. However, for $dt=0.3$, we find that $\Delta_d=0.074$, $\Delta_s=0.007$ and $\Delta_\pi=0.032$. More results for $\delta=1/8$ and various values of $dt$, including $K_{cdw}$, $\xi_s$ and $\xi_G$ are given in Table \ref{Stable}.

%==FigS3: Density-density correlation==
\begin{figure}
  \includegraphics[width=\linewidth]{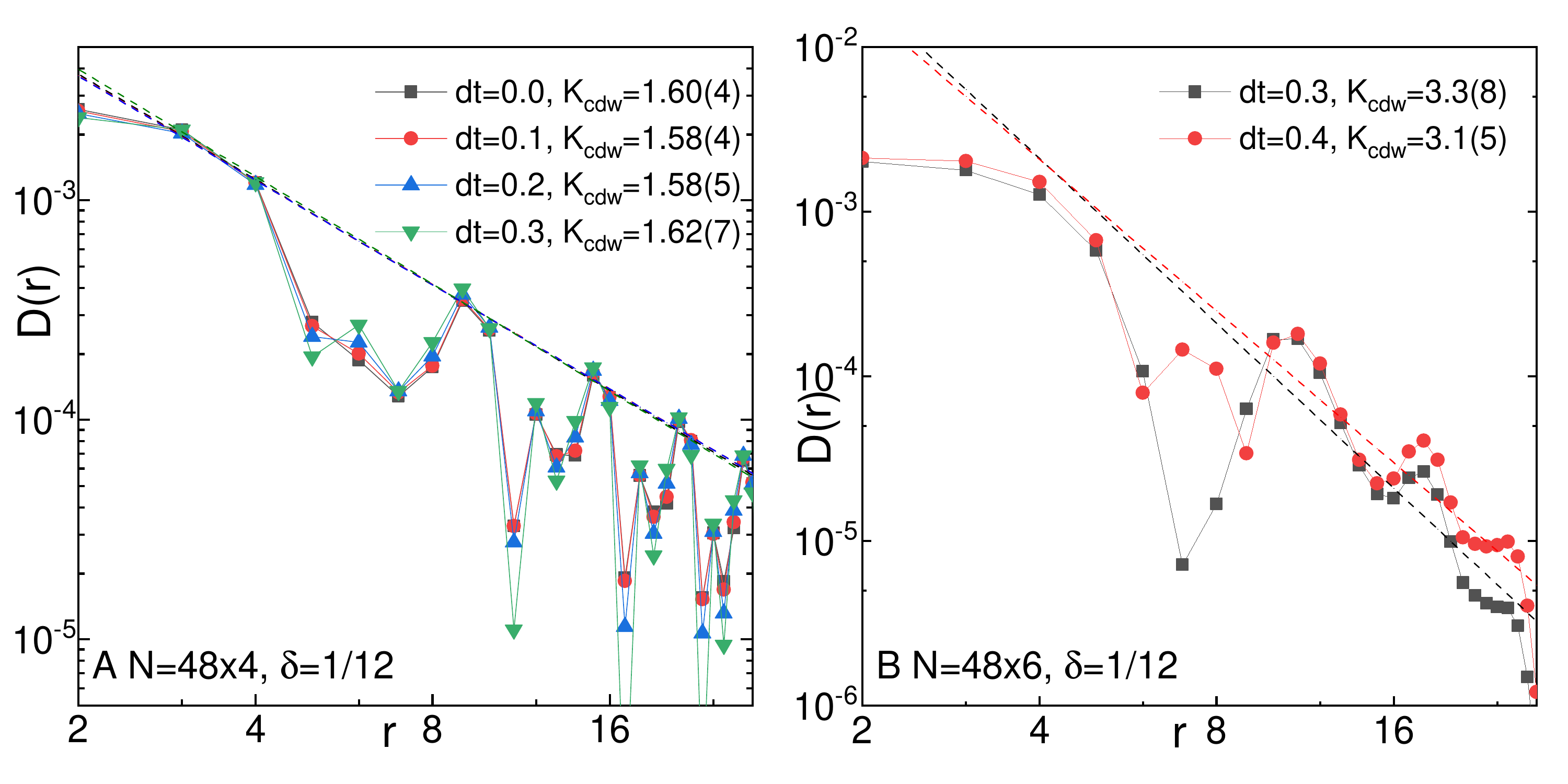}
  \caption{Charge density-density correlations. Charge density-density correlations $D(r)$ for (A) $N=48\times 4$ and (B) $N=48\times 6$ cylinders at $\delta=1/12$ doping level with different $dt$ on double-logarithmic scales, where $r$ is the distance between two sites in the $\hat{x}$ direction. The dashed lines denote a power-law fit $D(r)\sim r^{-K_{cdw}}$.
  }\label{FigS:DenCor}
\end{figure}

%==Extract K_c from charge density-density correlation==
\section{C. Charge density-density correlations}\label{SM:DenCor} %
The exponent $K_{cdw}$ was extracted from the $L_x$-dependence of the edge-induced CDW oscillations in the main text.  It can also be extracted from the charge density-density correlation, which is defined as $D(r)=\langle (\hat{n}(x_0)-\langle \hat{n}(x_0)\rangle)(\hat{n}(x_0+r)-\langle \hat{n}(x_0+r)\rangle)\rangle$. Here $x_0$ is the rung index of the reference site. Following a similar procedure as for $n(x)$ and $\Phi(r)$, the extrapolated $D(r)$ for a given cylinder in the limit $\epsilon=0$ or $m=\infty$ is obtained using a second-order polynomial function with the four data points of largest number of states.

As shown in Fig.\ref{FigS:DenCor}A and B for both $L_y=4$ and $L_y=6$ cylinders, $D(r)$ decays with a power-law at long distances, whose exponent $K_{cdw}$ was obtained by fitting the results using $D(r)\propto r^{-K_{cdw}}$. The extracted exponents for $L_y=4$ cylinders are $K_{cdw}
\approx 1.6$ for all studied values of $dt>0$ when $\delta=1/12$. For $L_y=6$ cylinders with $\delta=1/12$, $K_{cdw}=3.3(8)$ when $dt=0.3$ and $K_{cdw}=3.1(5)$ when $dt=0.4$. Note that $K_{cdw}$ extracted from $D(r)$ is slightly different from that extracted from the charge density oscillation $n(x)$ (see Table \ref{table}), which may be attributable to the boundary effect as well as the fact that the calculation of $D(r)$ is less accurate than $n(x)$ in the DMRG simulation. However, they are qualitatively consistent with each other and in all cases correspond  to $K_{cdw} > K_{sc}$.

\section{D. Symmetry considerations}
\label{symmetry}
The spatial symmetries of the striped Hubbard cylinder with $L_y$ an even integer are:  1)  Translation symmetry by any integer number of lattice constants in the x and any even number of lattice constants in the y direction.  2)  Reflection about a bond-centered line along the x or y axis and about a site-centered line along the y axis.  (Obviously, this implies inversion symmetry as well.) The model with $dt=0$ has, as, additional symmetries, translation by one lattice constant in the y direction and reflection about a site-centered line along the x axis.  In the limit that $dt=0$ and $L_y\to \infty$, there are additional $C_4$ rotational symmetries about both a site and a plaquette center, as well as reflection symmetry about a line along the $(1,1)$ direction.  In all cases, the model also has $SU(2)$ spin rotational symmetry and time-reversal symmetry.  Finally, for the half-filled band, the band-structure is particle-hole symmetric, leading\cite{YangZhang} to a formal SU(2) relating on site CDW and s-wave SC orders - although this presumably is unimportant for the repulsive $U$ case considered here.

In classifying possible spin singlet superconducting states, we focus on order parameters that are invariant under the translation symmetries of the model.  For finite $L_y$, there are two distinct irreducible representations (ireps.) of the point group which can be loosely referred to as   s-wave and d$_{xy}$ - with the latter being odd under the various reflections.  To the best of our knowledge, no evidence of $d_{xy}$ pairing  has been found for Hubbard ladders or cylinders have been seen in DMRG studies  to date, although a weak-coupling analysis of pairing in the Hubbard model\cite{raghu} suggests that such a state may arise for $\delta \gtrsim 0.5$. For $dt=0$ and $L_y\to \infty$, there are  additional possible pairing channels corresponding to the   $d_{x^2-y^2}$ and $g_{xy(x^2-y^2)}$  irreps.  However, all approaches to this problem lead to the conclusion that for $\delta$ not too large, the $d_{x^2-y^2}$ pairing is dominant.

All of this classification scheme is rendered somewhat less precise in 1D (i.e. for finite $L_y$) owing to the fact that there is only SC QLRO - no actual broken symmetries.  But we conjecture that even in this case -- from the asymptotic behavior represented in Eq.\ref{Eq:Ksc} -- we can use the same analysis to classify different forms of QLRO.

Naturally, it is also possible to consider SC states that spontaneously break translational symmetry, i.e. some form of pair-density-wave (PDW)\cite{pdw}.  The observation of $\pi$-pairing on the 4-leg cylinder can be thought of as an example of such a state -- and as such represents a symmetry distinct form of pairing when  $dt=0$.  However, as already mentioned, for $dt\neq 0$, this state is invariant under all the spatial symmetries, and so does not correspond to a distinct irrep.

\end{document}